\documentclass[amssymb,prb,aps,twocolumn,amsmath,showpacs]{revtex4}

\usepackage[dvips]{graphicx}
\usepackage{amstext}

\begin{document}

\title{Spin-memory loss at Co/Ru interfaces}
\author{Mazin A. Khasawneh}
\altaffiliation[Present address: ]{Laboratory for Physical
Sciences, 8050 Greenmead Drive, College Park, MD  20740.}
\affiliation{Department of Physics and Astronomy, Michigan State
University, East Lansing, Michigan 48824-2320, USA}
\author{Carolin Klose, W. P. Pratt, Jr., Norman O. Birge}
\email{birge@pa.msu.edu} \affiliation{Department of Physics and
Astronomy, Michigan State University, East Lansing, Michigan
48824-2320, USA}
\date{\today}

\begin{abstract}

We have determined the spin-memory-loss parameter,
$\delta_{Co/Ru}$, by measuring the transmission of spin-triplet
and spin-singlet Cooper pairs across Co/Ru interfaces in Josephson
junctions and by Current-Perpendicular-to-Plane Giant
Magnetoresistance (CPP-GMR) techniques.  The probability of
spin-memory loss at the Co/Ru interface is
$(1-exp(-\delta_{Co/Ru}))$.  From the CPP-MR, we obtain
$\delta_{Co/Ru} = 0.34^{+0.04}_{-0.02}$ that is in good agreement
with $\delta_{Co/Ru} = 0.35 \pm 0.08$ obtained from spin-triplet
transmission. For spin-singlet transmission, we have
$\delta_{Co/Ru} = 0.64 \pm 0.05$ that is different from that
obtained from CPP-GMR and spin-triplet transmission.  The source
of this difference is not understood.

\end{abstract}

 \maketitle

\section{Introduction}

Interest in spin-dependent transport in metals and semiconductors
has grown rapidly since the discovery of Giant Magnetoresistance
(GMR) in the late 1980's.  Optimization of the GMR and other
spin-dependent phenomena requires quantitative determination of
the parameters characterizing the spin-dependent transport in
ferromagnetic (F) and nonmagnetic (N) materials and at their
interfaces. Spin-dependent transport phenomena are widespread, and
are now known to produce exotic behavior in
superconducting/ferromagnetic (S/F) hybrid systems
\cite{Bergeret:01} as well.  The recent experimental demonstration
of induced spin-triplet pair correlations in S/F systems
\cite{Khaire:10} provides further impetus for understanding the
spin-dependent transport properties of F and N materials and
interfaces.

This paper focuses on the properties of the interface between Co
and Ru.  The Co/Ru/Co trilayer system exhibits strong oscillatory
exchange coupling between the two Co layers, and is known to be a
synthetic antiferromagnet (SAF) when the Ru thickness is in the
range 0.6 - 0.8 nm.\cite{Parkin:90}  In our recent work on S/F/S
Josephson junctions, we exploited the Co/Ru/Co SAF to form the
ferromagnetic (F) core of the junctions.  In the SAF, the
intrinsic magnetic flux due to the Co domains cancels in the two
Co layers, thereby allowing us to produce junctions with uniform
current density over large junction areas.\cite{Khasawneh:09} Such
junctions exhibit textbook-like Fraunhofer patterns when the
critical current is plotted vs. the magnetic field applied
perpendicular to the current direction -- in stark contrast to
S/F/S junctions with a single F layer in the place of the
SAF.\cite{Khasawneh:09}  When we placed additional ferromagnetic
layers (F') at either end of our junctions, to form S/F'/SAF/F'/S
structures, we observed long-range supercurrents due to the
generation of spin-triplet pair correlations in our
structures.\cite{Khaire:10}  Those supercurrents had been
predicted to appear in such systems in the presence of
non-collinear magnetizations between nearby ferromagnetic
layers.\cite{Bergeret:01,Houzet:07}  In our samples the
non-collinearity occurs between the F' and Co layers on either end
of the junctions.\cite{Khasawneh:11}

One issue that has not been resolved is the magnitude of the
critical current in our S/F'/SAF/F'/S Josephson junctions, with or
without the F' layers. The critical current in Josephson junctions
is often reported as the product of current times normal state
resistance, $I_cR_N$, because that product is normally independent
of the junction area. Our junctions exhibit $I_cR_N$ products of
order 1-10 $\mu V$ for very thin Co layers in the SAF; $I_cR_N$
then decreases exponentially with increasing Co
thickness.\cite{Khasawneh:09} Josephson junctions of the simpler
form S/Co/S have been fabricated and measured by Robinson
\textit{et al.}\cite{Robinson:06} Those workers fabricated
ultra-small junctions using a focused ion beam technique, and
reported $I_cR_N$ products as large as 1 mV, or about 100 times
larger than those of our samples.  Since the major difference
between the samples of Robinson \textit{et al.} and ours, aside
from lateral size, is the presence of the Ru layers in our
samples, an obvious candidate to explain our smaller values of
$I_cR_N$ is spin-flip scattering in the Ru or at the Co/Ru
interface.  One goal of this paper is to determine if such
scattering can indeed explain the critical current discrepancy
noted above.

Spin-flip and spin-orbit scattering are known to be important in
the context of GMR.\cite{Bass:07}  In the most obvious scenario,
spin-flip and spin-orbit scattering are sources of spin memory
loss in nonmagnetic (N) metals, which lead to the reduction of the
GMR signal in F/N/F devices.  In a less obvious scenario,
spin-flip scattering added intentionally at the outer edges of an
N/F/N/F/N device can \textit{increase} the GMR signal by limiting
the spatial extent over which the spin-up and spin-down electrons
carry current independently of each other.  The same effect
occurring in the F materials, however, leads to a reduction in the
GMR signal.  All of these effects can be understood quantitatively
using the Valet-Fert (V-F) equations to describe the
spin-dependent transport in the devices.\cite{Valet-Fert} Within
V-F theory, spin memory loss in a bulk metal is characterized by a
spin-memory length, $l_{sf}$.  The probability for an electron to
lose memory of its spin state while traversing a metal layer of
width $t$ is then $P=(1-\textrm{exp}(-t/l_{sf}))$. For an
interface between metals A and B, spin memory loss is
characterized by a dimensionless parameter $\delta_{A/B}$, with
the associated probability equal to
$P=(1-\textrm{exp}(-\delta_{A/B}))$.  The main goal of this paper
is the determination of $\delta_{Co/Ru}$ from a variety of
experiments, both in the context of superconducting systems and in
the context of GMR in non-superconducting systems.

\section{Suppression of Josephson supercurrent by Co/Ru interfaces}

In our recent work, we have measured the critical supercurrent,
$I_c$, in Josephson junctions of the form
Nb/Cu/F'/Cu/Co/Ru/Co/Cu/F'/Cu/Nb.\cite{Khaire:10,Khasawneh:11} The
inner Co/Ru/Co SAF possesses large exchange energy but very little
net magnetic flux.  The latter characteristic allows us to obtain
reliable estimates of $I_c$ from the measured Fraunhofer
patterns.\cite{Khasawneh:09}  In samples without F' layers, the
large exchange energy leads to a rapid decay of $I_c$ as the Co
thickness is increased.\cite{Khasawneh:09} This behavior is
well-understood,\cite{Buzdin:82} and is due to rapid dephasing of
the two electrons from the Cooper pair after they enter different
spin bands in the Co.\cite{Demler:95}  In samples with certain
specific F' layers, $I_c$ practically does not decrease with
increasing Co thickness,\cite{Khaire:10,Khasawneh:11} which is a
sign that the supercurrent is being carried by spin-triplet rather
than spin-singlet pairs.\cite{Bergeret:05}  Spin-triplet pairs are
not present in the original Nb superconductor, but they are
produced when the magnetization of the F' layer is non-collinear
with that of the nearest Co
layer.\cite{Houzet:07,Volkov:10,Trifunovic:10} The largest
spin-triplet supercurrent has been obtained with F' being a
4-nm-thick layer of PdNi alloy or a 2-nm-thick Ni layer. If the
total Co thickness is kept fixed at 20 nm, $I_c$ is enhanced by
more than two orders of magnitude by inserting either of those F'
layers.\cite{Khasawneh:11}

In this work we are interested in how the presence of Co/Ru
interfaces affects the magnitude of $I_c$.  Since the supercurrent
can be carried either by spin-singlet or spin-triplet pairs, there
are two parts to this question.  To address how Co/Ru interfaces
affect spin-singlet supercurrent, we have fabricated samples
without F' layers, and containing varying numbers of Ru layers.
The total Co thickness is kept fixed at 8 nm -- large enough to
allow subdivision into up to four sub-layers but small enough so
as not to suppress $I_c$ below our measurement sensitivity. The Ru
layers are always 0.6 nm thick, to optimize antiferromagnetic
coupling between the Co layers on either side. For $N=1$, the
central SAF is of the form Co(4)/Ru/Co(4); for $N=2$ it is
Co(2)/Ru/Co(4)/Ru/Co(2); and for $N=3$ it is
Co(2)/Ru/Co(2)/Ru/Co(2)/Ru/Co(2), where all thicknesses are in nm.
This design keeps the net magnetic flux as close to zero as
possible for each value of $N$.  To address how Co/Ru interfaces
affect spin-triplet supercurrent, we have fabricated samples with
4-nm thick PdNi layers as the F' layers. In these samples the
total Co thickness is kept fixed at 20 nm, which is enough to
suppress the spin-singlet supercurrent by at least two orders of
magnitude relative to the spin-triplet
supercurrent.\cite{Khaire:10} The ratios of the Co-layer
thicknesses are the same as before: for $N=1$, the central SAF is
of the form Co(10)/Ru/Co(10); for $N=2$ it is
Co(5)/Ru/Co(10)/Ru/Co(5); and for $N=3$ it is
Co(5)/Ru/Co(5)/Ru/Co(5)/Ru/Co(5).

\begin{figure}[tbh]
\begin{center}
\includegraphics[width=3.0in]{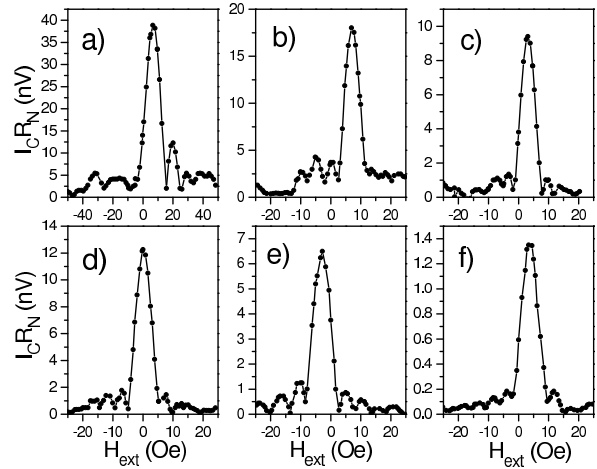}
\end{center}
\caption{Critical current $I_c$ times normal-state resistance
$R_N$ \textit{vs.} applied magnetic field ("Fraunhofer patterns")
for 6 Josephson junctions with (top row) or without (bottom row)
4-nm PdNi as F' layers (see text). The number $N$ of Ru layers is
a) $N=1$; b) $N=2$; c) $N=3$; d) $N=1$; e) $N=2$; f) $N=3$. All
data are from 20-$\mu$m diameter pillars except those in a), which
are from a 10-$\mu$m diameter pillar. }\label{RuFraunhofers}
\end{figure}

Raw data of $I_cR_N$ vs. applied magnetic field from
representative samples of all six types are shown in Fig. 1. The
quality of the Fraunhofer patterns is good for all six. The
central peaks are shifted from zero by only a few Oersteds,
indicating good flux cancellation -- i.e. antiferromagnetic
alignment of adjacent Co layer magnetizations.

\begin{figure}[tbh]
\begin{center}
\includegraphics[width=2.2in]{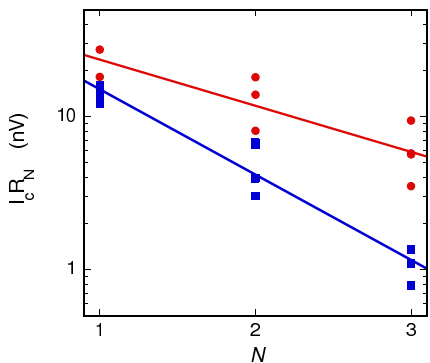}
\end{center}
\caption{(color online).  Product of critical current $I_c$ times
normal-state resistance $R_N$ \textit{vs.} number $N$ of Ru layers
in the Josephson junctions.  Red circles represent samples with
4-nm thick PdNi F' layers and with total Co thickness of 20 nm,
carrying spin-triplet supercurrent.  Blue squares represent
samples without F' layers and with total Co thickness of 8 nm,
carrying spin-singlet supercurrent. The lines are least-squares
fits discussed in the text.}\label{IcRN_vs_NRu}
\end{figure}

The dependence of $I_cR_N$ on the number of Ru layers is plotted
in Fig. 2, both for samples with (red circles) and without (blue
squares) F' layers.  In both cases the critical current decreases
with increasing number of Ru layers, but surprisingly, the rate of
decrease is different in the two cases.  Since each Ru layer
introduces two additional Co/Ru interfaces, we have fit each set
of data to an exponential decay of the form $I_cR_N \propto
exp(-2N\delta_{Co/Ru})$.  For the samples without F' layers, in
which we expect the supercurrent to be carried entirely by
spin-singlet electron pairs, the value of $\delta_{Co/Ru}$
obtained from the fit is $0.64 \pm 0.05$.  For the samples with F'
layers and with the thicker Co, in which the supercurrent is
carried almost entirely by spin-triplet pairs, the value of
$\delta_{Co/Ru}$ obtained from the fit is $0.35 \pm 0.08$.  We do
not understand why spin-singlet pairs appear to be suppressed more
than spin-triplet pairs at Co/Ru interfaces.

A number of theoretical works have discussed the effect of
spin-dependent, spin-flip, or spin-orbit scattering on the
critical current of S/F/S Josephson junctions in various
regimes.\cite{Demler:95,Oh:00,Bergeret:03,Faure:06,Houzet:05,Gusakova:06,Kashuba:07,Bergeret:07}
These works address scattering in the bulk of the materials,
rather than at interfaces, and most of them address the
spin-singlet rather than the spin-triplet supercurrent.  Ref.
\onlinecite{Bergeret:03} is an exception in that it addresses the
effect of spin-orbit scattering in the bulk of the F material on
both the spin-singlet and spin-triplet supercurrent in S/F/S
Josephson junctions.  The authors conclude that, for moderate
spin-orbit scattering, the spin-triplet component is more
sensitive than the spin-singlet component to the spin-orbit
interaction.  Thus it appears unlikely that spin-orbit scattering
is responsible for our experimental observations. There is also a
growing literature on spin-dependent boundary conditions at S/F
interfaces.\cite{Cottet:09}  To our knowledge, neither that
literature nor the works cited above provide microscopic
calculations of spin-memory loss at interfaces, and how such
spin-memory loss affects the spin-singlet and spin-triplet
supercurrent.

The main conclusion from the results presented in this section is
that each additional Ru layer causes only a mild suppression of
the critical supercurrent, by factors exp$(-2\times 0.64) = 0.28$
or exp$(-2\times 0.35) = 0.50$, for the singlet and triplet
supercurrents, respectively. This mild suppression is not enough
to explain the much larger difference between the values of
$I_cR_N$ observed in our large-area samples and those observed in
the much smaller samples studied by Robinson \textit{et
al.}\cite{Robinson:06}

In the next section we discuss an entirely independent way of
obtaining $\delta_{Co/Ru}$, this time by looking at the
propagation of spin-polarized electrons in the normal state --
without any superconductivity.

\section{Using Giant Magnetoresistance to estimate $\delta_{Co/Ru}$}

\subsection{Important CPP parameters}

Before we present the details of the sample structure to determine
$\delta_{Co/Ru}$, we define and quantify the important
current-perpendicular-to-plane (CPP) parameters that determine the
GMR and how the CPP transport is modeled.

For CPP diffusive transport in
ferromagnetic-/nonferromagnetic-metal (F/N) multilayers, the
following bulk and interface parameters are important.  In the
bulk of F, one has the resistivities $\rho_F^{\uparrow}$ and
$\rho_F^{\downarrow}$ that can be combined to give
$\rho_F^*=(\rho_F^{\uparrow}+\rho_F^{\downarrow})/4$ and
spin-asymmetry parameter
$\beta_F=(\rho_F^{\downarrow}-\rho_F^{\uparrow})/(\rho_F^{\downarrow}+\rho_F^{\uparrow})$.
The arrows ($\uparrow$) and ($\downarrow$) correspond to the
electron moment being parallel or antiparallel to the moment of F,
respectively. For F/N interfaces, one has the interface
resistances $AR_{F/N}^{\uparrow}$ and $AR_{F/N}^{\downarrow}$ that
combine to give
$AR_{F/N}^*=(AR_{F/N}^{\uparrow}+AR_{F/N}^{\downarrow})/4$ and
spin-asymmetry parameter
$\gamma_{F/N}^*=(AR_{F/N}^{\downarrow}-AR_{F/N}^{\uparrow})/(AR_{F/N}^{\downarrow}+AR_{F/N}^{\uparrow})$.
 For the simple case of no
electron-spin flipping in the multilayer, a two current series
resistor model (2CSR) can be used to analyze the MR
behavior.\cite{Lee:93,Valet-Fert} For the more general case of
spin flipping in the bulk of the layers and at the interfaces, the
Valet-Fert (VF) model\cite{Valet-Fert} must be solved numerically
to extract important parameters such as Co/Ru.

Our samples contain Nb, Cu, Py (= Permalloy $\approx$
Ni$_{0.8}$Fe$_{0.2}$), Co, Ru and FeMn.  Our own prior
studies\cite{Dassonneville:10,Fierz:90,Bass:99} give the following
parameters for these metals: $\rho_{FeMn}=875 \pm 50$ n$\Omega$m;
$AR_{Nb/FeMn}=1.0 \pm 0.6$ f$\Omega$m$^2$; $AR_{FeMn/Py}=1.0 \pm
0.4$ f$\Omega$m$^2$; $\rho_{Py}=123 \pm 40$ n$\Omega$m;
$\beta_{Py}=0.76 \pm 0.07$; $l_{sf}^{Py}= 5.5 \pm 1$ nm;
$AR_{Py/Cu}^*= 0.50 \pm 0.04$ f$\Omega$m$^2$; $\gamma_{Py/Cu}=0.7
\pm 0.1$. $\rho_{Co}=60 \pm 4$ n$\Omega$m; $\beta_{Co}=0.46 \pm
0.05$; $\gamma_{Co/Cu}=0.75 \pm 0.04$; $AR_{Co/Cu}^*=0.52 \pm
0.02$ f$\Omega$m$^2$; $\delta_{Co/Cu}=0.33_{-0.08}^{+0.03}$;
$l_{sf}^{Co}=60 \pm 20$ nm \cite{Piraux:98,Reilly:98}; $\rho_{Cu}
= 5 \pm 1$ n$\Omega$m; and from ref. \onlinecite{Bass:07},
$l_{sf}^{Cu}> 1000$ nm; $\rho_{Ru}=95$ n$\Omega$m and $l_{sf}^{Ru}
\approx 14$ nm.\cite{Eid:02} Prior CPP preliminary studies
indicated that $\gamma_{Co/Ru} \approx -0.2$ and $AR_{Co/Ru}^*
\approx 0.50$ f$\Omega$m$^2$.\cite{Eid:02} With a
Co(3nm)/Ru(0.6nm) multilayer, we obtained a similar value of
$AR_{Co/Ru}^*=0.60 \pm 0.03$ f$\Omega$m$^2$.\cite{Ahn:08} We will
refine $AR_{Co/Ru}^*$ and $\gamma_{Co/Ru}$ later in Section III.C.

\subsection{Sample Structures}

Two kinds of samples are employed to determine $\delta_{Co/Ru}$,
using CPP-GMR at 4.2K.  A CPP double exchange biased spin valve
(DEBSV) is used for both structures:\cite{Dassonneville:10}
Nb(150)/Cu(10)/FeMn(8)/Py(6)/Cu(10)/$X_i$/Cu(10)/
Py(6)/FeMn(8)/Cu(10)/Nb(150), where the thicknesses are in nm and
$X_i$ represents the inner sets of layers of the two samples
labelled with $i=1$ or 2.  The two Py layers are pinned by
exchange-bias coupling to the FeMn layers, so that their magnetic
moments reverse together at a much higher field $H$ than is needed
to reverse the overall moment of $X_i$. Also $t_{Cu}$=10 nm is
thick enough to exchange-decouple $X_i$ from the Py layers.  To
achieve uniform current flow in the CPP geometry, the multilayers
are sandwiched between $\sim 1.1$-mm wide, crossed Nb strips,
which superconduct at our measuring temperature of 4.2 K. We find
the overlap area $A \approx 1.2$ mm$^2$ through which the CPP
current flows by measuring the width of each Nb strip with a
Dektak profilometer. The intrinsic quantity for these measurements
is AR where R is the CPP resistance.  Our sputtering system,
sample preparation, and measuring techniques are described in ref.
\onlinecite{Lee:95}.

$X_1$ has the following structure: [Co(3)/Ru(1.4)/]$_n$Co(3) where
$n$ ranges from 0 to 8.  The 1.4-nm-thickness of the Ru is chosen
to cause parallel exchange coupling between the Co layers so that
the magnetizations of the Co layers switch as a single unit when a
modest in-plane magnetic field is reversed.  The CPP-GMR then
results from reversal of the moment of $X_1$ from parallel (P) to
anti-parallel (AP) to the common direction of the moments of the
two Py layers.  We measure $A\Delta R=AR^\mathrm{AP} - AR^\mathrm{P}$ and see how
it changes with $n$, a behavior that depends upon
$\delta_{Co/Ru}$. One of us used an identical sample structure
with Ru replaced by Cu to determine
$\delta_{Co/Cu}$.\cite{Dassonneville:10}

$X_2$ has the following structure:
[Co(1.5)/Ru(0.6)/]$_m$[Co(3)/Cu(1.4)/Co(3)/] [Ru(0.6)/Co(1.5)]$_m$
where $m$ ranges from 0 to 3.  The 1.4-nm-thick Cu layer in the
middle of the structure exchange couples the moments of the two
adjacent Co layers parallel.  The 0.6-nm-thick Ru layers couple
the adjacent Co-layer moments in an antiparallel state.  For
example, with $m = 1$, the Co(1.5) layers are antiparallel to the
nearest Co(3) layer, and the 1.5-nm-thickness of the outer Co
layers ensures that the overall moment of the multilayer is
parallel to that of the Co(3) layers.  Thus this $X_2$ system will
switch as a unit when a modest magnetic field is reversed.  As
will be explained later in Section III.D., this $X_2$ geometry
shows a more sensitive dependence of $A\Delta R$ on
$\delta_{Co/Ru}$ for $m = 1$ and 3.

\subsection{Refinement of $AR_{Co/Ru}^*$ and $\gamma_{Co/Ru}$}

\begin{figure}[tbh]
\begin{center}
\includegraphics[width=2.6in]{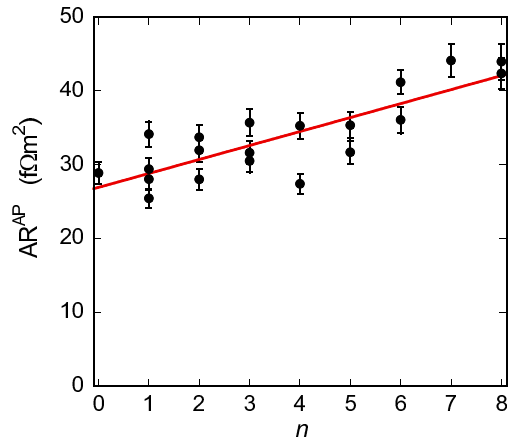}
\end{center}
\caption{$AR^\mathrm{AP}$ \textit{vs.} $n$ for the $X_1$ samples.  The line
is a linear least-squares fit to the data. }\label{Main}
\end{figure}

For $X_1$ samples, the slope from a plot of $AR^\mathrm{AP}$ vs $n$ can be
used to determine $AR_{Co/Ru}^*$.  The 2CSR model predicts that
slope=$\rho_{Co}^* \cdot 3 \textrm{nm} + \rho_{Ru} \cdot 1.4
\textrm{nm} +2AR_{Co/Ru}^*$. Fig. 3 shows a plot of $AR^\mathrm{AP}$ vs $n$.
The least-squares linear fit has a slope of $1.89 \pm 0.30$
f$\Omega$m$^2$ that gives $AR_{Co/Ru}^*=0.77 \pm 0.15$
f$\Omega$m$^2$.  This value of $AR_{Co/Ru}^*$ just agrees within
mutual independent uncertainties with $AR_{Co/Ru}^*=0.60 \pm 0.03$
f$\Omega$m$^2$  for $t_{Ru}=0.6$ nm.\cite{Ahn:08}  Because the
higher interface resistance may be associated with a more
completely-formed interface for $t_{Ru}=1.4$ nm, we will use the
$X_1$ value of $AR_{Co/Ru}^*$ for $X_1$-related calculations, and
we use the lower value of $AR_{Co/Ru}^*$ for the $X_2$ samples
where $t_{Ru} = 0.6$ nm.

To refine $\gamma_{Co/Ru}$, we revisit the preliminary analysis in
ref. \onlinecite{Eid:02}. In Fig. 3 of that publication, $A\Delta
R$ is plotted \textit{vs.} $t_{Co}$ for a multilayer of the form:
[Py(6)/Cu(20)/Ru(2)/Co($t_{Co}$)/Ru(2)/Cu(20)/Py(6)]$_6$.  If
$\gamma_{Co/Ru}$ is negative, there will be value of $t_{Co}$
where positive spin asymmetry in bulk of the Co cancels the
negative spin asymmetry of the two Co/Ru interfaces, and $A\Delta
R$=0. The 2CSR model predicts that

\begin{equation}\label{ADeltaR_X2}
A\Delta R \propto \beta_{Co} \rho_{Co}^* t_{Co} + \gamma_{Co/Ru}
2AR_{Co/Ru}^*
\end{equation}

A linear fit to $A\Delta R$ \textit{vs.} $t_{Co}$ gives
$t_{Co}=5.2 \pm 0.3$ nm at the point where $A\Delta R=0$.  Using
the $X_1$ value of $AR_{Co/Ru}^*$ (since $t_{Ru}=2$ nm here) and
the known values of other parameters (except for
$\gamma_{Co/Ru}$), we obtain $\gamma_{Co/Ru}=- 0.12 \pm 0.03$.
This value of $\gamma_{Co/Ru}$ will be used in the analysis of our
data.

\begin{table*}[tbh]
\begin{tabular}
[c]{|c|c|}\hline $m$ & Magnetic structure for $X_2$ sample
\\\hline
0 & Cu/$\Uparrow$/Cu/$\Uparrow$/Cu \\
1 & Cu/$\downarrow$/Ru/$\Uparrow$/Cu/$\Uparrow$/Ru/$\downarrow$/Cu \\
2 & Cu/$\uparrow$/Ru/$\downarrow$/Ru/$\Uparrow$/Cu/$\Uparrow$/Ru/$\downarrow$/Ru/$\uparrow$/Cu \\
3 &
Cu/$\downarrow$/Ru/$\uparrow$/Ru/$\downarrow$/Ru/$\Uparrow$/Cu/$\Uparrow$/Ru/$\downarrow$/Ru/$\uparrow$/Ru/$\downarrow$/Cu\\\hline
\end{tabular}
\newline \caption{Alignment of Co(3) ($\Uparrow$) layer moments with respect to those of the Co(1.5) ($\uparrow,\downarrow$) layers in the $X_2$ samples.}
\label{tableMoments}
\end{table*}

\subsection{Expected behavior of $A\Delta R$}

To simplify the data analysis that will come later, we present
here the expected behavior of $A\Delta R$ for both types of
samples using the V-F model. To determine the effect of
$\gamma_{Co/Ru}$, we will vary $\gamma_{Co/Ru}$ between the
extremes of 0 and -0.12. Also we know that $\delta_{Co/Cu} =
0.33_{-0.08}^{+0.03}$;\cite{Dassonneville:10} and, anticipating
our result for $\delta_{Co/Ru}$, here we set, temporarily,
$\delta_{Co/Ru}=\delta_{Co/Cu}$.

\begin{figure}[tbh]
\begin{center}
\includegraphics[width=2.4in]{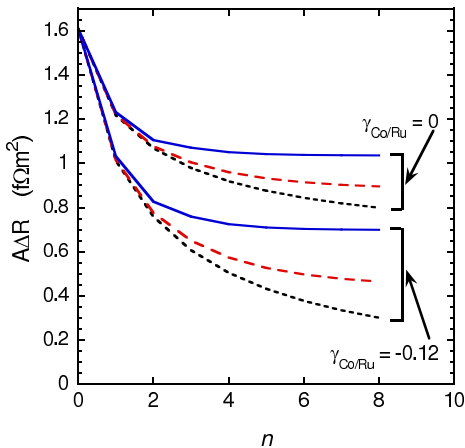}
\end{center}
\caption{(color online). $A\Delta R$ \textit{vs.} $n$ for $X_1$
samples with two values of $\gamma_{Co/Ru}$. The curves are
explained in text.  The transition from dashed to solid curves
show the effects of turning on finite spin flipping at the Co/Ru
interfaces. }\label{PrattFig1}
\end{figure}

For the $X_1$ samples, Fig. 4 shows the anticipated behavior of
$A\Delta R$ \textit{vs.} $n$ for $\gamma_{Co/Ru}$=0 and - 0.12.
Here $A\Delta R$ decreases with increasing $n$.  In contrast, the
Co/Cu system showed $A\Delta R$ increasing with
$n$.\cite{Dassonneville:10} This difference is due the positive,
much larger value of $\gamma_{Co/Cu}=+ 0.75$, in comparison to
$\gamma_{Co/Ru}$.

The dotted curves are for spin flipping at the Co/Cu interfaces
only with no spin flipping elsewhere.  The dashed curves show what
happens when bulk spin flipping in Co and Ru is added.  Finally,
the solid curves exhibit the additional effects of having
$\delta_{Co/Ru}=0.33$.  The transition from the dashed to solid
curves at large $n$ indicates that, in principle, the effects of
finite $\delta_{Co/Ru}$ should be observable in the data. However,
the significant dependence of $A\Delta R$ on $\gamma_{Co/Ru}$ will
likely complicate the extraction of $\delta_{Co/Ru}$ from the
data.

\begin{figure}[tbh]
\begin{center}
\includegraphics[width=2.4in]{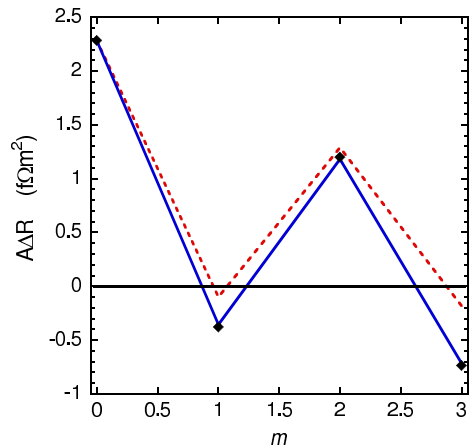}
\end{center}
\caption{(color online).  $A \Delta R$ \textit{vs.} $m$ for $X_2$
samples.  The lines and diamond symbols are explained in the text.  The transition
from the dotted to solid lines shows the effects of turning on
finite spin flipping at the Co/Ru interfaces.}\label{Pratt_Fig_2}
\end{figure}

For the $X_2$ samples, Fig. 5 shows the expected behavior of
$A\Delta R$ \textit{vs.} $m$ using the V-F model.  The lines are
for $\gamma_{Co/Ru}=- 0.12$. The dotted lines are for spin
flipping only at Co/Cu interfaces and in the bulk of the Co and Ru
layers. These dotted lines are hardly changed if the bulk spin
flipping is omitted. The solid lines show what happens to $A\Delta
R$ when $\delta_{Co/Ru}=0.33$.  As anticipated, $A\Delta R$ is
most sensitive to $\delta_{Co/Ru}$ for $m=1$ and 3, as explained
below. The diamond symbols represent the case where
$\gamma_{Co/Ru}=0$.  In contrast to Fig. 3, here $A\Delta R$ is
not sensitive to $\gamma_{Co/Ru}$, and this lack of sensitivity
will make a determination of $\delta_{Co/Ru}$ more robust.  For
example, if the calculated value of $A\Delta R$ for
$\gamma_{Co/Ru}=- 0.12$ and $\delta_{Co/Ru}=0.33$ is held fixed as
$\gamma_{Co/Ru}$ is set to zero, $\delta_{Co/Ru}$ only decreases
to 0.31.

Including the Cu layers between the $X_2$ insert and the two Py
layers, the sample structure is
Cu(10)/[Co(1.5)/Ru(0.6)/]$_m$[Co(3)/Cu(1.4)/Co(3)/]
[Ru(0.6)/Co(1.5)]$_m$/Cu(10). This structure is designed to make
the major contribution to $A\Delta R$ of the large asymmetry of
the Co/Cu interfaces ($\gamma_{Co/Cu}=+ 0.75$) sensitive to
$\delta_{Co/Ru}$ for $m=1$ and 3.

Table 1 shows how the magnetic moments of the two Co(3)
($\Uparrow$) layers are aligned with respect to the Co(1.5)
($\uparrow,\downarrow$) layers.

For no spin flipping anywhere, the 2CSR model predicts the
following behaviors.  (a) For $m = 0$ and 2, the moments of the
outer Co(1.5)/Cu interfaces are parallel to those of the inner
Co(3)/Cu interfaces, giving a large positive $A\Delta R$ as seen
in Fig. 5.   (b) For $m = 1$ and 3, the outer Co(1.5)/Cu
interfaces are antiparallel to those of the inner Co(3)/Cu
interfaces, giving $A\Delta R \approx 0$, as seen Fig. 5.  The
contributions of Co/Ru regions to $A\Delta R$ are included in
items (a) and (b), but we clarify next how the 2CSR model applies
to these Co/Ru regions. For $m = 1 - 3$, the two antiparallel
Co/Ru interfaces on each side of a given $t_{Ru}=0.6$ nm layer
together give no contribution to the overall spin asymmetry of the
$X_2$ layer and thus do not contribute to $A\Delta R$.  For
example, the contributions to $A\Delta R$ of the
$\uparrow$/Ru/$\downarrow$ interfaces are $+\gamma_{Co/Ru}
AR_{Co/Ru}^*$ for the left interface and $-\gamma_{Co/Ru}
AR_{Co/Ru}^*$ for the right.  Since this cancelation is
independent of the value of $\gamma_{Co/Ru}$, $A\Delta R$ will not
depend on $\gamma_{Co/Ru}$, at least for the case of no spin
flipping anywhere.  Thus the very weak dependence of $A\Delta R$
upon $\gamma_{Co/Ru}$ shown in Fig. 5 is plausible.  With $m = 1$
and 3, the overall bulk-Co contribution to $A\Delta R$ is
equivalent to two Co layers ($\uparrow$) with 1.5 nm thickness,
giving in terms of the 2CSR model that $A\Delta R=+2(\beta_{Co}
\rho_{Co}^{*}t_{Co})=0.1$ f$\Omega$m$^2$. This positive bulk contribution
to $A\Delta R$ is small compared to that for $m = 0$ and 2,
justifying the statement that $A\Delta R \approx 0$ for $m = 1$
and 3.

For finite $\delta_{Co/Ru}$ with $m = 1$ and 3, the spin flipping
at the Co/Ru interfaces causes the two outer $\downarrow$/Cu
interfaces to make a more significant contribution to $A\Delta R$
than the central two $\Uparrow$/Cu interfaces.  Crudely, the two
inner $\Uparrow$/Cu interfaces are becoming "isolated" by spin
flipping from the two outside Py layers.  Thus we have $A\Delta R
< 0$, as seen in Fig. 5.  For $m = 3$, $A\Delta R$ is even more
negative because there are more Co/Ru interfaces between the inner
and outer Co/Cu interfaces. Hopefully these negative values of
$A\Delta R$ will be seen in the data.

\section{Giant Magnetoresistance data}

\subsection{$X_1$ raw data}

Raw magnetoresistance data for a representative selection of $X_1$
samples with $n=1$ and $n=6$ are shown in Fig. 6.  The outer Py
layers are pinned in the negative direction in the figure.  Hence
at negative applied fields $H$, the samples are in the P (low
resistance) state, and at positive $H$, the samples are in the AP
(high resistance) state. The hysteresis is due to the coercive
field of the inner Co/Ru multilayer, which switches sharply as a
single block due to the ferromagnetic exchange coupling induced by
the 1.4-nm thick Ru layers.  The flat parts of the graphs outside
the hysteresis region allow precise determination of $\Delta R$.

\begin{figure}[tbh]
\begin{center}
\includegraphics[width=3.2in]{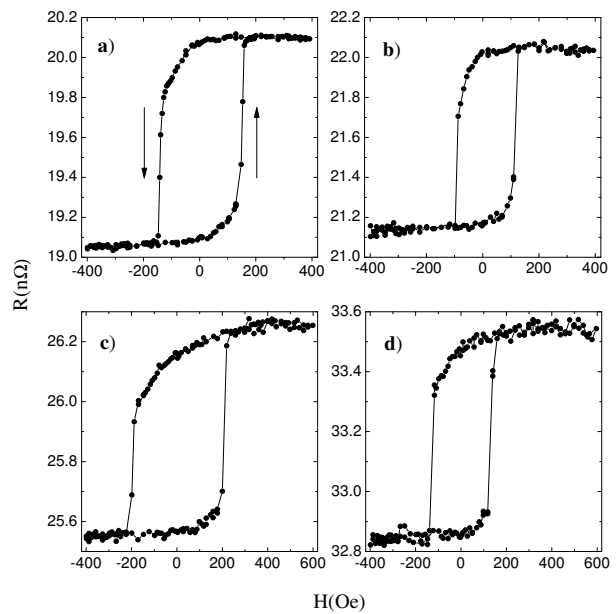}
\end{center}
\caption{Magnetoresistance data for two $X_1$ samples with $n=1$
(panels \textbf{a} and \textbf{b}) and two with $n=6$ (panels
\textbf{c} and \textbf{d}).}\label{MazinRaw1}
\end{figure}

\subsection{$X_2$ raw data}

Raw magnetoresistance data for a representative selection of $X_2$
samples with various values of $m$ are shown in Fig. 7. Again, the
outer Py layers are pinned in the negative direction.  A sample
with $m=0$, shown in panel a), exhibits clear switching of the
central Co/Cu/Co trilayer, as expected, at $\sim \pm$ 200 Oe.  At negative $H$, the sample
is in the P (low resistance) state, while at positive $H$, the
sample is in the AP (high resistance) state.  Note that $X_2$
samples with $m=0$ are similar to $X_1$ samples with $n=1$, except
that the former have Cu rather than Ru separating the two central
Co layers. As a result, they have much larger values of $A\Delta
R$ because $\gamma_{Co/Cu}$ is large and positive whereas
$\gamma_{Co/Ru}$ is small and negative.

\begin{figure}[tbh]
\begin{center}
\includegraphics[width=3.2in]{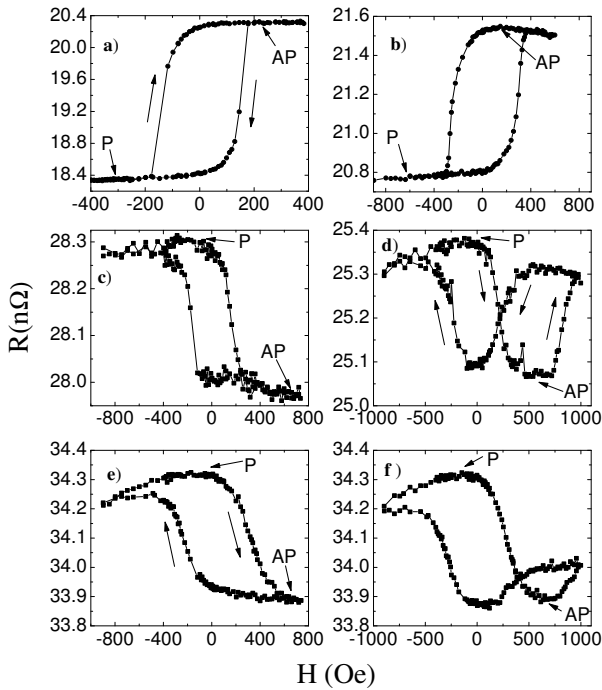}
\end{center}
\caption{Magnetoresistance data for several $X_2$ samples:
\textbf{a} $m=0$; \textbf{b} $m=2$; \textbf{c} and \textbf{d}
$m=1$ (the latter shows data with $H$ taken to larger values, with
depinning of the Py layers when $H > 750$ Oe); \textbf{e} and
\textbf{f} $m=3$ (the latter panel shows unpinning of the Py layers
when $H > 700$ Oe).  Note that when the Py layers are unpinned by a
large positive value of $H$, repinning occurs only at a
significantly lower value of $H$. Labels P and AP point to the
regions that determine $R^\mathrm{P}$ and $R^\mathrm{AP}$,
respectively.}\label{MazinRaw2}
\end{figure}

A sample with $m=2$, shown in panel b), exhibits similar behavior,
since for any even value of $m$ the outer Co layers are parallel to
the central Co layers.  The resistance data in the AP state (at $H
> 400$ Oe) are not quite as flat in this sample as in the $m=0$
sample, so $R^\mathrm{AP}$ was determined near $H=150$ where $R$
is a maximum.  Note that the switching field here is approximately
twice that for $m=0$ in panel a). So at these higher fields we may
be seeing the effects of unpinning of the Py layers that would
cause $R$ to decrease with increasing field.

The situation is different for samples with odd values of $m$, as
shown in panels c) and e) of Fig. 7 for lower fields and panels d)
and f) for higher fields.  These four panels show that the
magnetoresistance is negative, as expected if $\delta_{Co/Ru}$ is
finite.  $R^\mathrm{P}$ is determined from the data near $H$=0
because $R$ decreases as $H$ becomes more negative.  In this
decreasing field, those outer thin-Co layers that are antiparallel
to the central Co layers are tending to rotate parallel to the
central Co and Py layers, moving toward a "global" parallel state
for all of the Co layers.  This will decrease $R$.  In fact, for
$H << 0$, we estimate that $R$ will decrease by $\sim1.3$
n$\Omega$ for the sample in panel d).  The decrease in R seen in
panel d) as $H$ goes from 0 to $\sim - 1000$Oe is much smaller
than this predicted extreme case, so the misalignment of the
antiparallel outer Co layers is small but still makes a
significant contribution to the magnetoresistance.  Thus is
important to determine $R^\mathrm{P}$ near $H=0$. $R^\mathrm{AP}$
is determined in the low-slope region near $H$=+600 Oe, just below
the onset of Py-layer unpinning for $H > 750$. As discussed for $H
< 0$, increasing $+H$ will cause those outer thin-Co layers that
are antiparallel to the central Co layers to rotate more parallel
to the central layers and less parallel to the Py layers. Thus $R$
will tend to increase.  This means that the actual value of
$R^\mathrm{AP}$ is likely to be less than $R^\mathrm{AP}$ directly
determined from data for $H$=$\sim$ +600 Oe. Thus the magnitude of
$A\Delta R$ is underestimated.  To make this correction to
$R_\mathrm{AP}$, one could take the change in $R$ as $H$ varies
from $\sim$ - 600 Oe to 0 Oe and subtract this from the nominal
value of $R^\mathrm{AP}$.  We will analyze the $A\Delta R$ data
without this correction and then ask what happens to
$\delta_{Co/Ru}$ when this correction is made to the $m=1$ data.

For $m=3$, making such corrections to $R^\mathrm{AP}$ is more
difficult.  As shown in panels e) and f) of Fig. 7, a larger value
of $H$ is needed to saturate the magnetization of the
[Co(1.5)/Ru(0.6)/]$_3$[Co(3)/Cu(1.4)/Co(3)/] [Ru(0.6)/Co(1.5)]$_3$
free layer than for the $m=1$ samples.  Thus the "plateau" region
where $R^\mathrm{AP}$ is evaluated contains a competition between
an approach to saturation of the Co-containing multilayer and the
unpinning of the Py layers.  Thus it is likely that
$R^\mathrm{AP}$ is over estimated, causing the magnitude of
$A\Delta R$ to be underestimated.  Hence, the experimental values
of $A\Delta R$ are less reliable even if the above-mentioned
corrections of the previous paragraph were applied to the $m=3$
samples.

\section{Data analysis and discussion}

\subsection{$A\Delta R$ data for $X_2$ samples}

As we showed in Fig. 5, $A\Delta R$ for $X_2$ samples is not
sensitive to $\gamma_{Co/Ru}$.  So we analyze the $X_2$ samples
first to establish a value $\delta_{Co/Ru}$ that we can compare
with that from the Josephson junction studies in Section II.  Note
that the Josephson junctions and the $X_2$ samples employ the same
$t_{Ru}$=0.6 nm thickness, while the $X_1$ samples have
$t_{Ru}$=1.4 nm.

\begin{figure}[tbh]
\begin{center}
\includegraphics[width=2.4in]{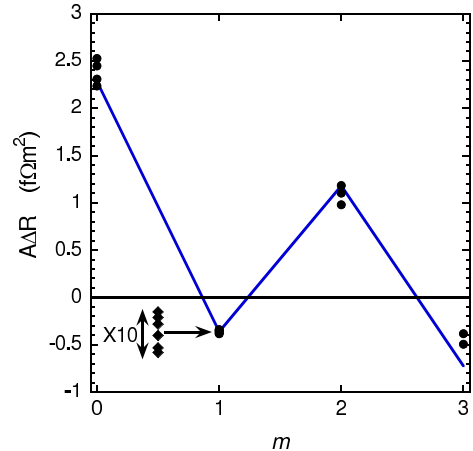}
\end{center}
\caption{(color online). $A \Delta R$ \textit{vs.} $m$ for all of the $X_2$
samples.  The fitting lines are described in the text. The solid
circles are the data points, and solid diamonds are a 10X vertical
expansion of the $m = 1$ data for clarity.}\label{MazinDataFit1}
\end{figure}

Fig. 8 shows $A\Delta R$ \textit{vs.} $m$ for the $X_2$ samples.
The solid lines represent the fit of V-F model to the $m = 1$ data
only, using $\gamma_{Co/Ru}=- 0.12$ and the other parameters
presented in Section III.A, and obtaining $\delta_{Co/Ru}=0.34$.
The diamond symbols show a detail of the six data points for
$m=1$, where the ordinate is expanded by a factor of ten about the
average value of the data. For the $m$ = 0 samples where there are
no Co/Ru interfaces present, the calculated value of $A \Delta R$
agrees well with the average value of data within mutual
uncertainties. This agreement strengthens the argument that the
parameters tabulated in Section III.A are relevant to our $X_2$
samples.  For $m = 2$, the calculated value of $A\Delta R$ also
agrees well with data. For $m = 3$, the experimental value of
$A\Delta R$ does not agree very well with its calculated value, as
anticipated in the last paragraph of Section IV.B.

If one applies the $R^\mathrm{AP}$ correction outlined in Section
IV.B to the $m=1$ data, $\delta_{Co/Ru}$ only increases to 0.35.
Varying $\gamma_{Co/Ru}$ by its $\pm 0.03$ uncertainty only
contributes a $\pm 0.004$ uncertainty to $\delta_{Co/Ru}$, as
expected from the discussion of Fig. 5.  Including the uncertainty
associated with $\delta_{Co/Cu} = 0.33_{-0.08}^{+0.03}$, we obtain
a final value of $\delta_{Co/Ru} = 0.34_{-0.02}^{+0.04}$.
Interestingly, this value of $\delta_{Co/Ru}$ agrees with those
obtained for Co/Cu and Co/Ni interfaces: $\delta_{Co/Cu} =
0.33_{-0.08}^{+0.03}$ and $\delta_{Co/Ni} = 0.33 \pm
0.06$.\cite{Dassonneville:10, Nguyen:10}

\subsection{$A\Delta R$ data for $X_1$ samples}

\begin{figure}[tbh]
\begin{center}
\includegraphics[width=2.4in]{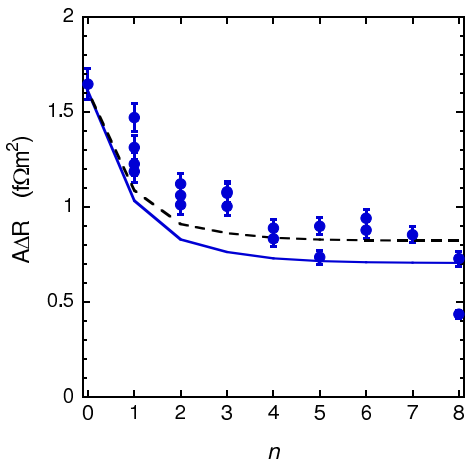}
\end{center}
\caption{(color online). $A \Delta R$ \textit{vs.} $n$ for all of the $X_1$
samples.  The fitting curves are described in the
text.}\label{MazinDataFit1}
\end{figure}

Fig. 9 shows $A\Delta R$ \textit{vs.} $n$ for the $X_1$ samples.
The solid curve indicates the expected behavior for the V-F model
employing the parameters that were used in the V-F fits to the
data in Fig. 8.  Although the overall drop in $A\Delta R$ with
increasing $n$ is reproduced, the fit is not very good especially
for small $n$. The dashed curve shows the V-F model fit when
$\gamma_{Co/Ru} = -0.09$ and $\delta_{Co/Ru} = 0.38$ are used, the
extreme values allowed by their uncertainties.  Most of the rise
in $A\Delta R$ is due to the increase in $\gamma_{Co/Ru}$, as
expected from the discussion concerning Fig. 4.  While this dashed
curve fits the larger $n$ data pretty well, it still does not fit
the low-$n$ data for reasons that are not understood. So we rely
on the fits to the data of the $X_2$ samples to determine
$\delta_{Co/Ru}$.  However, the $AR^\mathrm{AP}$ \textit{vs.} $n$
data for these $X_1$ samples with $t_{Ru}$=1.4 nm was useful in
determining $AR_{Co/Ru}^*$ so that the value of $\gamma_{Co/Ru}$
could be refined (see Section III.C).

\section{Conclusions}

The interfacial spin-memory loss parameter, $\delta_{Co/Ru}$, has
been determined in two ways: measuring the transmission of
spin-triplet and spin-singlet Cooper pairs across Co/Ru interfaces
in Josephson junctions, and using Current-Perpendicular-to-Plane
Giant Magnetoresistance techniques (CPP-MR).  For spin-triplet
transmission, we obtain $\delta_{Co/Ru} = 0.35 \pm 0.08$ in
comparison to $\delta_{Co/Ru} = 0.34_{-0.02}^{+0.04}$ from CPP-MR
measurements.  These two values of $\delta_{Co/Ru}$ are in
excellent agreement.  They also agree with $\delta_{F/N}$ values
obtained for Co/Cu and Co/Ni interfaces: $\delta_{Co/Cu} =
0.33_{-0.08}^{+0.03}$ and $\delta_{Co/Ni} = 0.33 \pm
0.06$.\cite{Dassonneville:10, Nguyen:10}  It is hoped that this
agreement will stimulate more theoretical work to establish the
source(s) of spin-memory loss at F/N interfaces.  The most likely
contributions include spin-orbit and interfacial spin-disorder
scattering.\cite{Fert-Lee:96, Tang-Wang:08}  Also we refined an
earlier estimate of the Co/Ru interfacial scattering
asymmetry\cite{Eid:02} and obtained $\gamma_{Co/Ru}=- 0.12 \pm
0.03$.  For spin-singlet transmission across the Co/Ru interface,
we obtained $\delta_{Co/Ru} = 0.64 \pm 0.05$ that is about a
factor of two larger than $\delta_{Co/Ru}$ from spin-triplet and
CPP-MR measurements.  This factor-of-two enhancement in
$\delta_{Co/Ru}$ is not understood and will hopefully encourage
further theoretical work.

Acknowledgments:  We thank S. Bergeret for helpful discussions, R.
Loloee and B. Bi for technical assistance, and use of the W.M.
Keck Microfabrication Facility.  This work was supported by the
U.S. Department of Energy under grant DE-FG02-06ER46341.

\end{document}